\documentclass[final,authoryear,1p,times]{elsarticle}

\usepackage{graphics}
\usepackage{graphicx}
\usepackage{epsfig}

\usepackage{amssymb}

\usepackage{lineno}

\usepackage{longtable}

\journal{Icarus}

\begin{document}

\begin{frontmatter}


\title{Planetary and satellite three body mean motion resonances.}

\author{Tabar\'e Gallardo\corref{cor1}}
\ead{gallardo@fisica.edu.uy}
\author{Leonardo Coito}
\author{Luciana Badano}

\cortext[cor1]{Corresponding author}

\address{Departamento de Astronom\'{i}a, Instituto de F\'{i}sica, Facultad
de Ciencias, Universidad de la Rep\'{u}blica, Igu\'{a} 4225, 11400 Montevideo, Uruguay}

\begin{abstract}

We propose a semianalytical method to compute the strengths on each of the three massive bodies participating in
a three body mean motion resonance (3BR).
 Applying this method we explore the dependence of the strength on the masses, the orbital parameters and the order of the resonance and we compare with previous studies. We confirm that for low eccentricity low inclination orbits zero order resonances are the strongest ones;
but for excited orbits higher order 3BRs become also dynamically relevant.
By means of numerical integrations and the construction of dynamical maps we check some of the predictions of the method.
We numerically explore the possibility of a planetary system to be trapped in a 3BR due to a migrating scenario. Our results suggest that capture in a chain of two body resonances is more probable than a capture in a pure 3BR. When a system is locked in a 3BR and one of the planets is forced to migrate the other two can react migrating in different directions.
We exemplify studying the case of the Galilean satellites where we show the relevance of the different resonances acting on the three innermost satellites.

\end{abstract}


\end{frontmatter}



\section{Introduction}
\label{intro}

One of the most prevalent dynamical phenomena observed in planetary systems  is orbital commensurability, or resonance.
Two body resonances (2BRs), extensively studied in orbital dynamics, occur when the ratio between the
mean motions, $n$, of two bodies can be written as a
fraction of 2 small integer numbers. They have proven to be very important in the architecture of the planetary systems \citep{fa14,baty15}.
A less common case of resonance ensues when the mean motions
of three bodies $P_0$, $P_1$ and $P_2$ verify
\begin{equation}\label{n}
   k_0n_0 +k_1n_1 +k_2n_2  \simeq 0
\end{equation}
being $k_i$ small integers, generating which is called a three body resonance (3BR).
In some cases, the 3BRs can be the consequence of a chain of two 2BRs as is the case of the
Galilean satellites studied since Laplace.
In fact, the three innermost Galilean satellites, Io, Europa and Ganymede, verify the 2BR relations $n_I - 2n_E \sim 0$ and  $n_E - 2n_G \sim 0$.
Subtracting both expressions  we obtain the 3BR $n_I - 3 n_E + 2n_G \sim 0$, called Laplacian resonance.
The resulting dynamics it is not a mere addition of the two 2BRs and the emerging 3BR generates a new complex dynamics.
The Laplacian
resonance is a paradigmatic case of a 3BR generated by the superposition or chains of two 2BRs.
On the other hand, there are also
3BRs that cannot be decomposed as chains of 2BRs and we will call them \textit{pure}. Thousands of asteroids
in pure 3BRs with Jupiter and Saturn can be found  in the Solar System \citep{ss13}.

A relevant parameter of the 3BRs is the order defined as $q=|k_0+k_1+k_2|$. It is known that
the lower the order the larger the dynamical effects of the resonance. That is why between the Galilean satellites the dominant 3BR is
$n_I - 3 n_E + 2n_G \sim 0$, and not
for example $n_I - n_E - 2n_G \sim 0$
which is of order 2 and obtained adding the 2BRs instead of subtracting them.
Note that the resonant condition (\ref{n}) can be written as
\begin{equation}\label{nn}
k_1(n_1-n_0) +k_2(n_2-n_0) + (k_0+k_1+k_2)n_0  \simeq 0
\end{equation}
which means that for zero order resonances, even in the case of pure 3BRs, the planets $P_1$ and $P_2$ are in a simple 2BR $k_1$:$k_2$ when
looked from the rotating frame of the planet $P_0$. No other 3BRs have this property which makes zero order 3BRs a special case.
Then, it is not surprising that zero order 3BRs have been deserved most the attention. They were studied for example by \citet{ak88} who obtained general formulae with applications in the asteroid belt and systems of satellites.
The case of Laplacian  resonance in the Galilean satellites has been intensely studied \citep{si75,fm79,ma91,sm97,ssm97,pl02,la09}.
Superposition or chains of 2BRs were also studied in the major Saturnian satellites \citep{cy10} and in extrasolar systems \citep{li11,ma13,bm13,ba15,pa15}.  \citet{qu14} focused on systems with close orbits with applications to the inner Uranian satellites, where it is remarked that 3BRs as consequence of superposition of first order 2BRs are the strongest ones. On the other hand,
pure 3BRs were studied for example by \citet{la84} for the specific case of the Uranian satellites and by \citet{nm99} where a complete planar theory was developed for the asteroidal, massless, case.
The situation among the outer planets of the Solar System was analyzed numerically by \citet{gu05,gu06}.
\citet{qu11} developed an analytical theory for general zero order resonances between three massive bodies in very close orbits while
\citet{ga14} developed a semianalytical method for estimation of the resonance's strength for pure 3BRs of any order for the asteroidal case assuming the perturbing planets in circular and coplanar orbits and the asteroid in an arbitrary orbit.
Finally, it is worth mention that \citet{sh15} suggested that  the satellites of Pluto,  Styx, Nix and Hydra, are driven by the zero order 3BR $3n_S-5n_N+2n_H \sim 0$.

\subsection{Looking for the disturbing function}
\label{look}

The dynamics of a system trapped in a 3BR  is determined by the resonant disturbing function, which its obtention
is not a trivial point.
The disturbing function for a 3BR emerges after a second averaging procedure applied on the
resulting expressions of a first averaging involving the mutual perturbations between the planets taken by pairs \citep{nm99}.
The final expression of the resonant disturbing function for planet $P_0$ assumed in the resonance
 given by Eq. (\ref{n})
 is a summatory  of  the type
\begin{equation}\label{term}
\mathcal{R} = k^2 m_1 m_2 \sum_{j} \mathcal{P}_j \cos (\sigma_j)
\end{equation}
where $k$ is the Gaussian constant and $m_1$ and $m_2$ the planetary masses,
with the critical angle
\begin{equation}\label{sigmaj}
\sigma_j = k_0 \lambda_0 + k_1 \lambda_1 +  k_2 \lambda_2 +  \gamma_j
\end{equation}
and
\begin{equation}\label{gamma}
\gamma_j =  k_3 \varpi_0 +  k_4 \varpi_1 +  k_5 \varpi_2 + k_6 \Omega_0  + k_7 \Omega_1 + k_8 \Omega_2
\end{equation}
being $\lambda$, $\varpi$ and $\Omega$ the mean longitudes, longitudes of the perihelia and longitudes of the nodes respectively, $k_0,k_1,k_2$ are integers fixed by the resonance and the $k_{i>2}$  are arbitrary integers but verifying the d'Alembert condition
\begin{equation}\label{dala}
\sum_{i=0}^{8}k_i=0
\end{equation}
$\mathcal{P}_j$ is a polynomial  function depending on the eccentricities and inclinations which its lowest order term is
\begin{equation}\label{termpoly}
C e_0^{|k_3|} e_1^{|k_4|} e_2^{|k_5|}  \sin(i_0)^{|k_6|} \sin(i_1)^{|k_7|} \sin(i_2)^{|k_8|}
\end{equation}
 The calculation of the coefficients $C$ is a very laborious task that must be done case by case and it is so challenging that only the planar case was studied by analytical methods and consequently there are not expansions involving $\sin(i_i)$ published up to now.
An example of this development can be found in \citet{go12} where an expansion for a specific 3BR in an extrasolar planar system is obtained. The expansion
given by Eq. (\ref{term}) implies that for a given resonance there are several $\sigma_j$ contributing to the
resonant motion.
 Each $\sigma_j$ generates specific dynamical effects and the joint action of all $\sigma_j$ is called multiplet.
Nevertheless, the expansion (\ref{term}) can be reduced to a few terms when the eccentricities and inclinations are very small.
In particular,
when $e_1=e_2=i_1=i_2=0$
the lowest order non null terms for $\mathcal{P}_j$
are those with $k_4=k_5=k_7=k_8=0$:
\begin{equation}\label{nonu}
C e_0^{|k_3|} \sin(i_0)^{|k_6|} \cos (k_0 \lambda_0 + k_1 \lambda_1 +  k_2 \lambda_2 +  k_3 \varpi_0 + k_6 \Omega_0)
\end{equation}
from which can be deduced that for three coplanar orbits ($i_0=0$) the only non null terms are those with $k_6=0$, and consequently the lowest order term in the expansion is proportional to $e_0^{q}$,  where $q=|k_3|$.
This explain why the lower the order the stronger the resonance.
In case that $e_0=0$ but with
$i_0\neq 0$ the non null terms are those with $k_3=0$ which result proportional to  $\sin(i_0)^{q}$ instead, where $q=|k_6|$. But, as we explain below,
if
$|k_6|$ is odd the resulting principal term of the expansion is  proportional to  $\sin(i_0)^{2q}$.
 Note that for coplanar circular orbits all terms are null except for zero order resonances because in this special case the principal terms are independent of $e_i, i_i$.

To avoid the difficulties of the analytical methods \citet{ga14} proposed a semianalytical method for the estimation of the strength of a resonance on a massless particle in an arbitrary orbit under the
effect of two perturbing planets in circular coplanar orbits. The method, which is essentially an estimation of the amplitude of the disturbing function factorized by an arbitrary constant coefficient,  was applied to minor bodies captured in 3BRs with the planets of the Solar System. In the present work, in section \ref{depen} we extend the method to a system of three massive bodies with arbitrary orbits and we apply it to an hypothetical planetary system in order to analyze the dependence of the strengths on the orbital parameters.
In section \ref{numeric} we explore by numerical methods some of the properties of the resonances that our method predicts and we apply the method to
the case of the Galilean satellites. The conclusions are presented in section \ref{conclu}.

\section{Strength for planetary three body resonances and its dependence with the parameters}
\label{depen}

Strictly, 3BRs between three planets $P_0$, $P_1$ and $P_2$ with elements ($a_i$, $e_i$, $i_i$, $\Omega_i$, $\varpi_i$)
and masses $m_0$, $m_1$ and $m_2$ around a star of mass $M$
occur when a particular critical angle given by Eq. (\ref{sigmaj})
is oscillating over time.
In this work  we call $p=|k_0| + |k_1| +|k_2| $
and we note as $k_0+k_1+k_2$ the resonance involving the three planets, where always $k_0 >0$.
We will not consider the case of 3BRs as result of superposition of 2BRs because the 2BRs
will override the dynamical effects of the 3BR we are trying to study, with the exception of systems with near zero eccentricity orbits.
We will consider the planets $P_1$ and $P_2$ at fixed semimajor axes $a_1 < a_2$
 and the third "test" planet $P_0$  with the semimajor axis defined by the resonant condition which can
 result in an internal, external o middle position with respect to $P_1$ and $P_2$.
 The approximate nominal location of the test planet $P_0$ assumed in the resonance
$k_0+k_1+k_2$ is deduced from Eq. (\ref{n}):
\begin{equation}\label{nominal}
    a_0^{-3/2} \simeq -\frac{k_1 \sqrt{(M+m_1)}}{k_0 \sqrt{(M+m_0)}}a_1^{-3/2} -\frac{k_2 \sqrt{(M+m_2)}}{k_0 \sqrt{(M+m_0)}}a_2^{-3/2}
\end{equation}
which must be positive otherwise the resonance does not exist.
In order to obtain a numerical estimation of the resonance's strength we extended the method given by \citet{ga14} to a system of three massive bodies
with arbitrary orbits.
The details of the method and the devised algorithm can be found in the Appendix. Essentially, this new method predicts different strengths called
$S_0, S_1, S_2$ for the three massive bodies, that means, each massive body feels the resonance in a different way.
Each $S$ is
related to the amplitude of the variations of $\mathcal{R}$ in Eq. (\ref{term}) caused by the cumulative effect of all involved terms. Then, the method cannot distinguish between the dynamical effects of each term of a multiplet for a given resonance, it only provides a global estimation.

\begin{figure}[h]
\resizebox{1.0\textwidth}{!}{\includegraphics{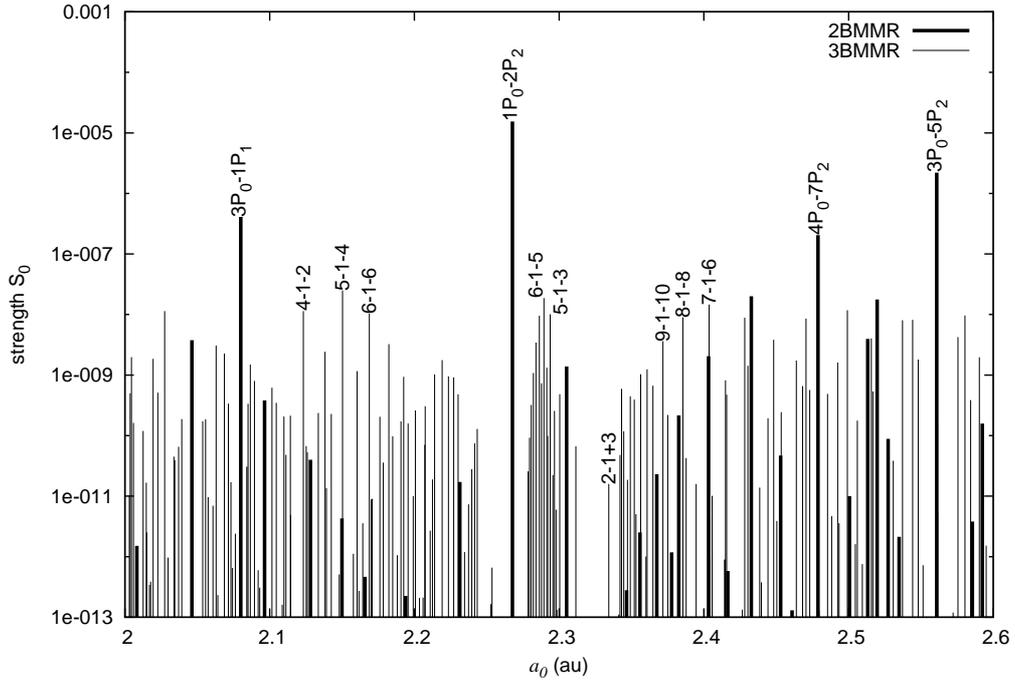}}
\caption{Location and strength of the main 3BRs (thin lines) and
 location and relative strengths of the 2BRs (thick lines) for the hypothetical working system with a planet $P_1$ at 1 au and planet $P_2$ at 3.6 au. The horizontal axis corresponds to the value of the semimajor axis $a_0$ of the test planet $P_0$ and the vertical axis corresponds to the strengths of the possible 3BRs.
The strengths were calculated assuming $e_i=0.05$, $i_i=1.0$ degree, $m_i=0.0001$ and the other elements taken from Table \ref{table0}.
Some two body and three body resonances are labeled. The 2BRs were plotted in a different scale than the 3BRs and
they are indicated only for reference.}
\label{atlas2y3}
\end{figure}

In order to test the algorithm and to explore the dependence of the strengths with the different parameters involved we applied it to an hypothetical planetary
 system with $m_1 = m_2 = 0.0001 M_{\odot}$, $a_1 = 1.0$ au, $a_2=3.6$ au around a star with 1 $M_{\odot}$ and we calculate all resonances with $q\leq 9$ and $p \leq 30$
between 2.0 au and 2.6 au, that means with the planet $P_0$ located in between and excluding close-encounter situations.
With the exception of section \ref{sectiongal}, in the examples presented along this paper $P_0$ is located between the other two planets,
but our method is valid for arbitrary positions of $P_0$ with respect to $P_1$ and $P_2$.
The
complete set of orbital parameters with their variation range used in our experiments can be found in Table \ref{table0}.
 Figure  \ref{atlas2y3} shows the main resonances in the interval where the strength of the 2BRs involving $P_0$ with planets $P_1$ or $P_2$ were calculated following the algorithm proposed in \citet{ga06} and the 3BRs were calculated with the algorithm proposed here. The set of 2BRs is not in the same scale of the set of 3BRs because they have different definitions. All codes can be downloaded from www.fisica.edu.uy/$\sim$gallardo/atlas.

\begin{table}
  \centering
 \begin{tabular}{  l   r    r  r  r r r  }
  \hline
 body         &    $a$ (au) &   $e$       &   $i (^{\circ})$     & $\Omega (^{\circ})$   & $\varpi (^{\circ})$  &     $m$ $(M_{\odot})$ \\
 \hline
   P$_0$      & $(2, 2.6)$  &  $(0, 0.3)$   &  $(0, 10)$   &    0       &    60     & $(0, 0.01)$ \\
   P$_1$      & 1.0       &   $(0, 0.1)$ & $(0, 10)$ &   120      &   180     &      $1\times 10^{-4}$ \\
   P$_2$      & 3.6       &   $(0, 0.1)$ & $(0, 10)$ &   240      &   300     &      $1\times 10^{-4}$  \\

  \hline	 								
\end{tabular}	 								
 \caption{Working example of an hypothetical planetary system with the range of variation of the orbital elements assumed in the calculations.
 The mass of the central star is 1 $M_{\odot}$.}\label{table0}			
\end{table}

\subsection{Effect of varying $m_0$ and the restricted case, $m_0 = 0$}

\begin{figure}[h]
\resizebox{1.0\textwidth}{!}{\includegraphics{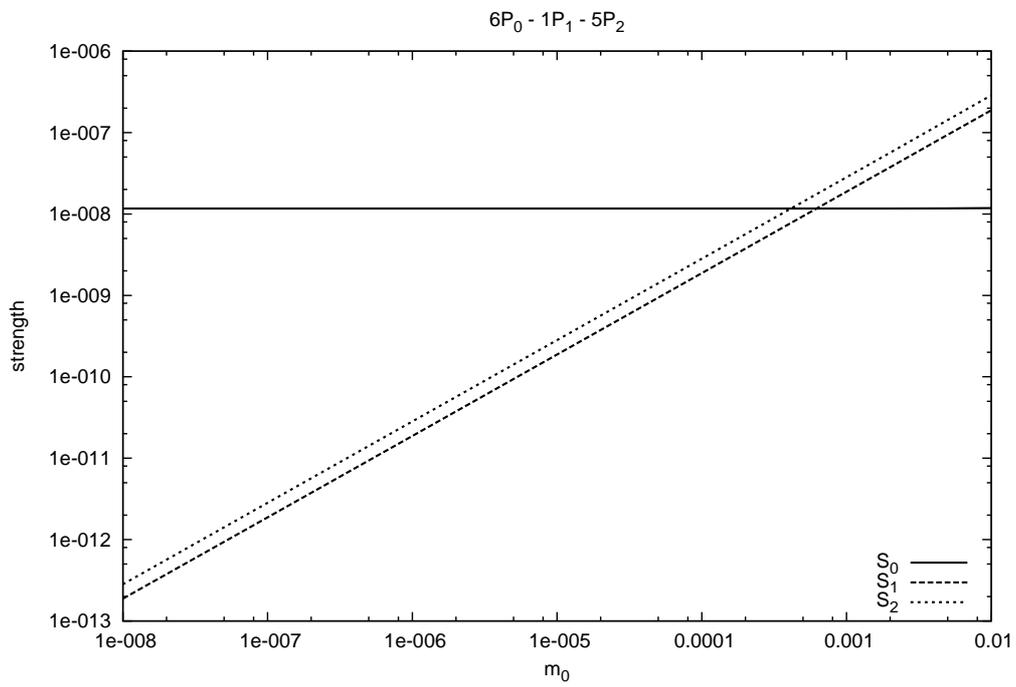}}
\caption{Strengths as function of $m_0$ for resonance $6-1-5$. The three planets are
assumed with $e_i=0.1$ and $i_i=0$. The planet having its mass varying is not affected by its own mass ($S_0$ is constant) but the other two planets have strengths
proportional to $m_0$. When the three masses are equal the greater strength is $S_0$, that means the planet in the middle. }
\label{stmass}
\end{figure}

To test the effect of the planetary mass on the strengths we choose the zero order resonance $6-1-5$ located at $a = 2.2894$ au and calculate the three strengths $S_0, S_1, S_2$ varying $m_0$. The results presented in Fig.  \ref{stmass} show that
when $m_0$ tends to zero $S_0$ is unaffected but $S_1,S_2$ tend to zero, that means the 3BR over
$P_0$ survive because is proportional to $m_1m_2$ but the other two planets tend to loose the resonance because their respective
strengths are factorized by $m_0$. This behaviour is similar for all resonances independently of the order.
For growing $m_0$, $S_0$ is unaffected but $S_1, S_2$ grow proportionally to $m_0$ and when $m_0$ is equal to the other masses, $S_0$
nevertheless is greater than $S_1$ and $S_2$. In general, for similar masses  the planet in the middle is the one with the greater dynamical effect, which is in agreement  with results obtained by \citet{qu11} for zero order resonances.
The fact that $S_0$ is independent of $m_0$ is not evident from the equations (\ref{R}) to  (\ref{sri}) but it is an evident result from
the analytical theories of 3BRs, see for example \citet{fm79} or \citet{qu11}. This concordance between numerical and analytical results gives support to our proposed algorithm, at least with respect to the role of the involved masses.

When considering the restricted case, $m_0 \rightarrow 0$, with P$_1$ and P$_2$ in circular and coplanar orbits it is easy to show that we reproduce the results
of the restricted case obtained by \citet{ga14}. That means $S_1 \sim S_2 \sim 0$ and it is clear a strong dependence of $S_0$ on the order $q$.
 For coplanar orbits $S_0 \propto e_0^{q}$ and for zero eccentricity orbits  $S_0 \propto \sin (i_0)^{q}$ for even $q$ and
  $S_0 \propto \sin (i_0)^{2q}$ for odd $q$ as in \citet{ga14}. This dependence on inclination is understood because by d'Alembert rules the inclinations only appear with even exponents in the development of the disturbing function.
  For $e_0=0$ the lowest order term is proportional to
\begin{equation}
\label{sin}
   \sin(i_0)^{|k_6|}  \cos (k_0 \lambda_0 + k_1 \lambda_1 +  k_2 \lambda_2  + k_6 \Omega_0 )
\end{equation}
and due to Eq. (\ref{dala}) we have $ |k_6| = |k_0 + k_1 + k_2| = q$, being $k_6$ even.
Then, if $q$ is odd the lowest order non null term is the next harmonic, which is
\begin{equation}
\label{sin2}
\sin(i_0)^{|2k_6|}  \cos (2k_0 \lambda_0 + 2k_1 \lambda_1 +  2k_2 \lambda_2  + 2k_6 \Omega_0 )
\end{equation}
and then   $S_0 \propto \sin (i_0)^{2q}$.

Now, we can apply the present method to the case of excited orbits for $P_1$ and $P_2$, not considered in \citet{ga14}. In this case we obtain that the
dependence with $q$ is not so clearly defined as in the case with circular planar orbits for $P_1$ and $P_2$ showed in \citet{ga14}. We can explain this result looking at the relevant terms
of the disturbing function. For the coplanar case with $e_1=e_2=0$ the only relevant term is the one factorized by $e_0^{q}$.  But, for non circular orbits
several terms factorized by $e_0^{l}e_1^{m}e_2^{n}$, with $l,m,n$ integers, contribute to the disturbing function and, if there is some mutual inclination,
terms depending on the inclinations will also appear. We will show this behaviour below for the non restricted case ($m_0 > 0$).

\subsection{Dependence with the resonance order $q$}

\begin{figure}[h]
\resizebox{1.0\textwidth}{!}{\includegraphics{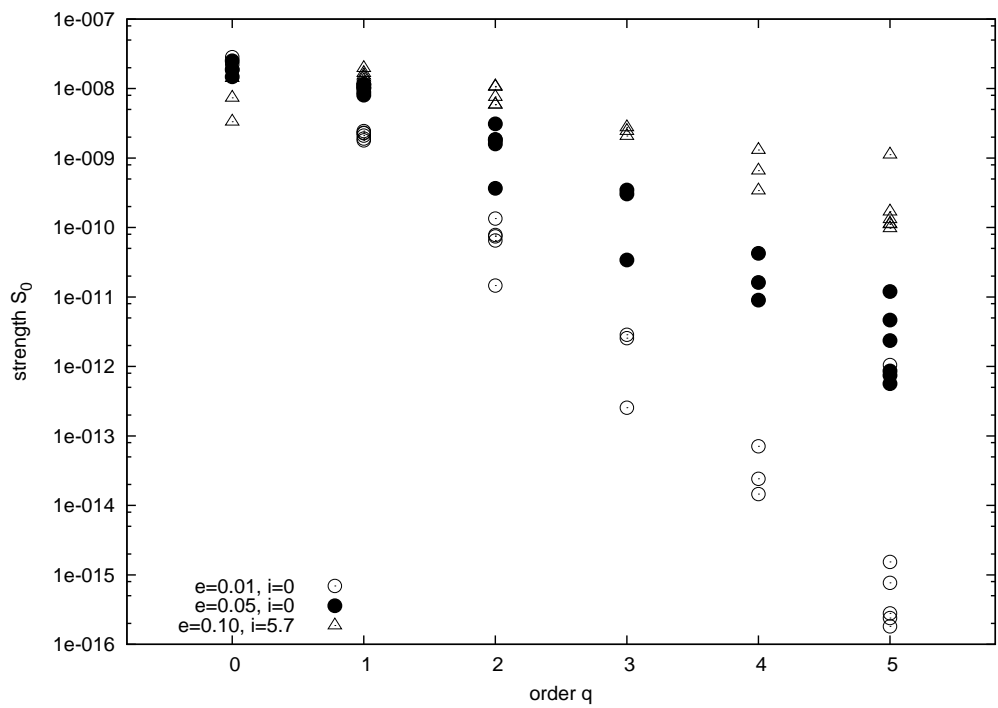}}
\caption{Strengths for planet P$_0$ of the lowest order 3BRs from Fig. \ref{atlas2y3}.
Strengths were calculated for
three different orbital states: coplanar and near zero eccentricities (open circles), coplanar and low-excited eccentricities (filled circles) and
dynamically excited orbits (triangles). The greater the dynamical excitation the lesser the dependence with the order $q$. }
\label{orderecc}
\end{figure}

\begin{figure}[h]
\resizebox{1.0\textwidth}{!}{\includegraphics{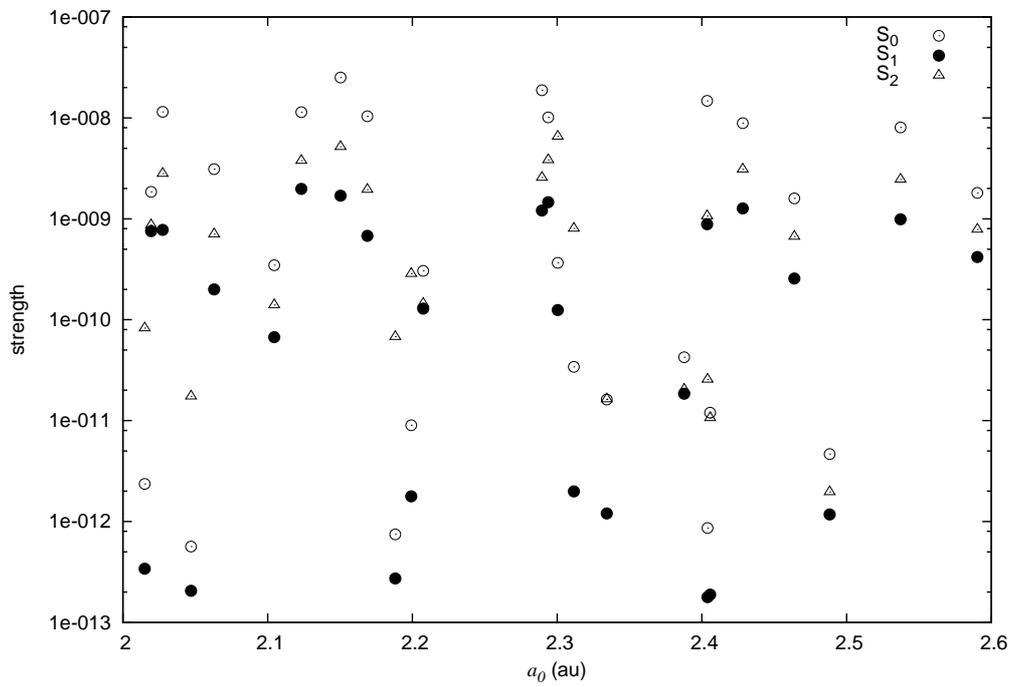}}
\caption{Strengths for the three planets for some  3BRs from Fig. \ref{orderecc} calculated for coplanar orbits with $e=0.05$. For each resonance defined by $a_0$ the strengths for each planet are showed.
In general it is verified $S_1 < S_2 < S_0$. }
\label{3strena}
\end{figure}

Fig.  \ref{orderecc} shows the strengths for planet $P_0$ assumed to be located in each of all resonances between 2.0 and 2.6 au with $p<15$ and $q\leq 5$ for three
different orbital configurations for the three planets: coplanar and almost circular orbits, coplanar and Jupiter-like eccentricity orbits and
dynamically excited orbits ($e=\sin(i)=0.1$). The dependence with the order is strong for near zero eccentricities
in contrast with
the excited
orbits where high order resonances are as important as the zero order.
Then, for near coplanar and circular orbits only zero order resonances are dynamically relevant, but
for excited orbital configurations high order resonances can also be dynamically relevant, a not surprising fact that it is well known for the case of 2BRs.
Fig.  \ref{orderecc} refers to the strength $S_0$ over planet $P_0$ located in between but we have also analysed  the strength $S_1$ over the innermost planet
and $S_2$ over the exterior planet. In Fig. \ref{3strena} we show the three $S_i$ for each resonance where we obtained systematically $S_1 < S_0$, that means, the inner planet experiences less dynamical effects
than the planet in the middle. For $q\leq 1$ we obtained also  $S_2 < S_0$ in agreement with \citet{qu11}, but for $q\geq 2$ the rule is not always verified.

\subsection{Dependence with eccentricity}

\begin{figure}[h]
\resizebox{1.0\textwidth}{!}{\includegraphics{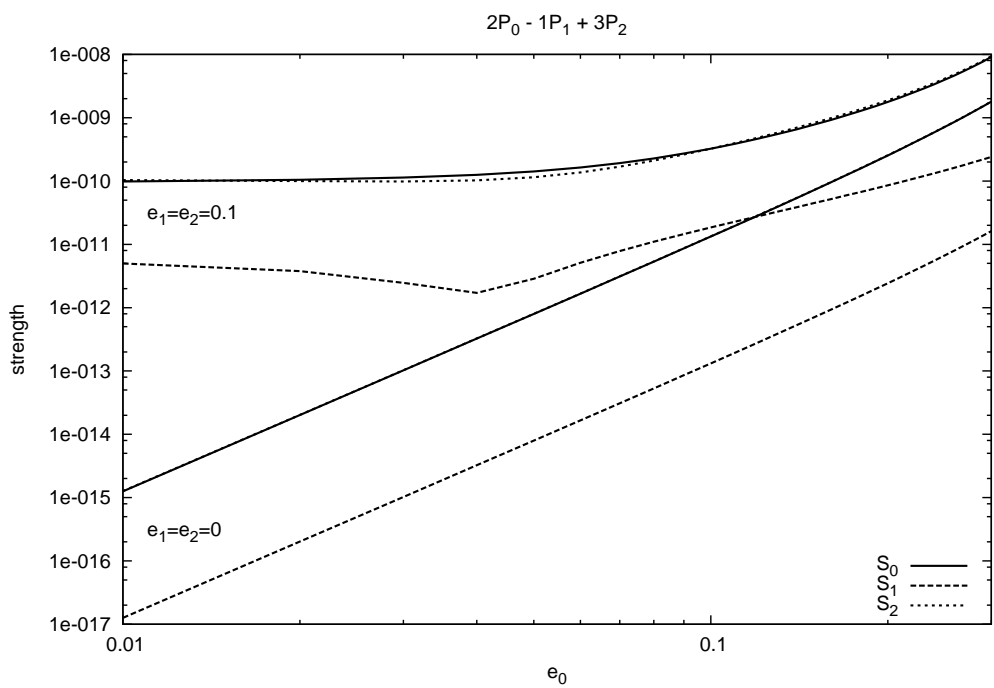}}
\caption{Strengths $S_0,S_1,S_2$ for the three planets in coplanar orbits as function of $e_0$ for the four order resonance $2-1+3$. Two cases are showed: $e_1=e_2=0$ in lower curves and the excited case $e_1=e_2=0.1$ in upper curves.
The dependence of $S_i$ with $e_0$ is mathematically very clear in the first case ($S_i \propto e_0^4$) but not in the excited one.
$S_0$ and $S_2$ are almost equal in this resonance.}
\label{2m13}
\end{figure}

Fig.  \ref{2m13} shows the strengths as function of $e_0$ for the four order resonance $2-1+3$ located at  $2.3343$ au  taking coplanar planets with $e_1=e_2=0$ in one case and with $e_1=e_2=0.1$ in other case.
In the first case the strengths $S_0$, $S_1$ and $S_2$ are proportional to $e_0^{q}$ being $q=4$ as expected for a four order resonance.
For the second case, when the planets are in eccentric orbits, the dependence with $e_0$ is not so clear mathematically.
The reason is that when the perturbing planets are eccentric there is not a unique term governing the
disturbing function as we have already explained, see for example \citet{nm99} and \citet{go12}. This behaviour for circular and excited orbits is similar for all non zero order resonances.

\subsection{Dependence with inclination}

\begin{figure}[h]
\resizebox{1.0\textwidth}{!}{\includegraphics{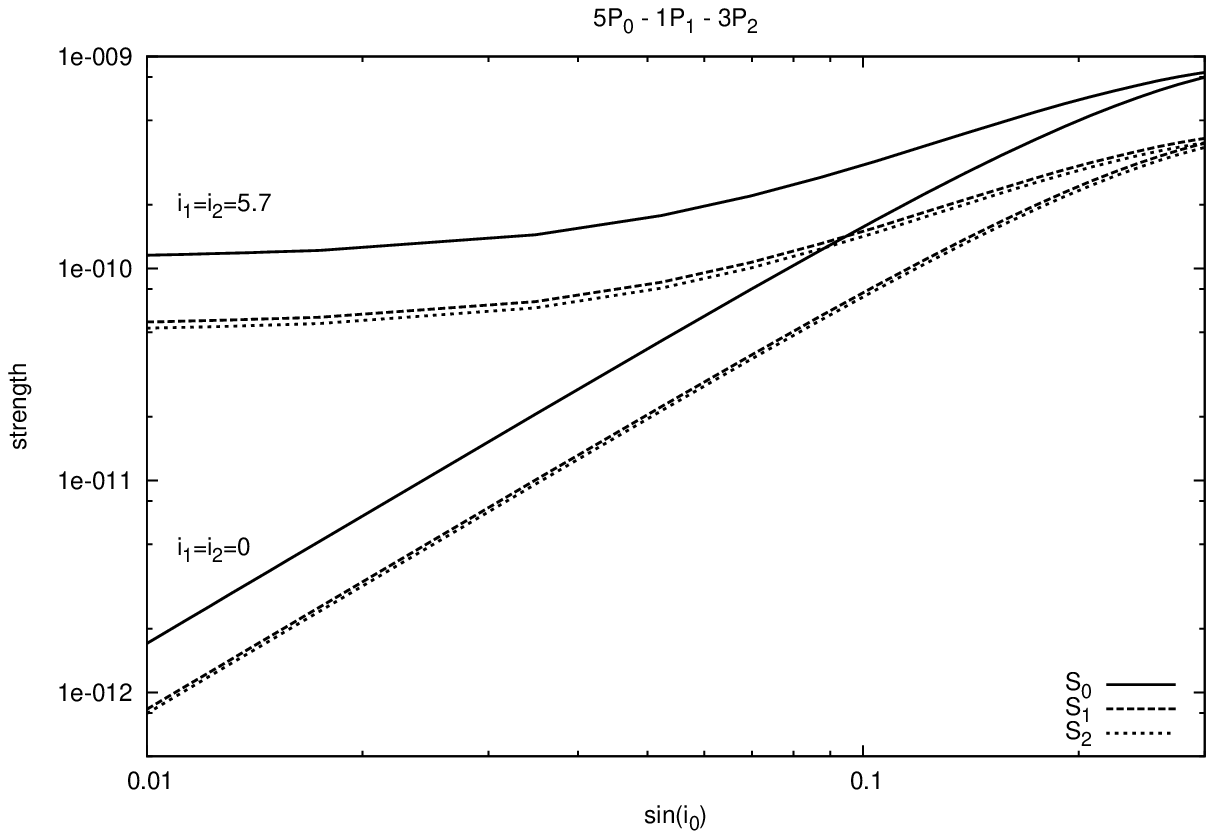}}
\caption{Strengths $S_0,S_1,S_2$ for the three planets assumed in circular orbits as function of $i_0$ for the first order resonance $5-1-3$. Two cases are showed: $i_1=i_2=0$ in lower curves and the inclined case $i_1=i_2=5.7$ in upper curves. In analogy with Fig. \ref{2m13}, the dependence of $S_i$ with $i_0$ is mathematically very clear in the first case ($S_i \propto \sin(i_0)^2$)  but not in the second one.}
\label{5m1m3}
\end{figure}

One of the advantages of the semianalytical method is that we can easily explore the dependence of
the resonances strengths with orbital inclinations.
Fig.  \ref{5m1m3} shows the strengths as function of $i_0$ for the first order resonance $5-1-3$ located at  $2.2939$ au  taking circular orbits with $i_1=i_2=0^{\circ}$ in one case and with $i_1=i_2=5.7^{\circ}$ in other case.
 In the first case, the strengths depend
on $(\sin i)^{2q}$ as expected for an odd-order resonance as  we have explained above.
 In analogy to the case of a planetary system with excited orbits, the circular but inclined cases show  resonance strengths with a not so well defined dependence with $i_0$ due to the several relevant terms in the disturbing function.
 This figure is probably the first one published showing the effect of the orbital inclination on a planetary 3BR.

\subsection{Zero order resonances}

\begin{figure}[h]
\resizebox{1.0\textwidth}{!}{\includegraphics{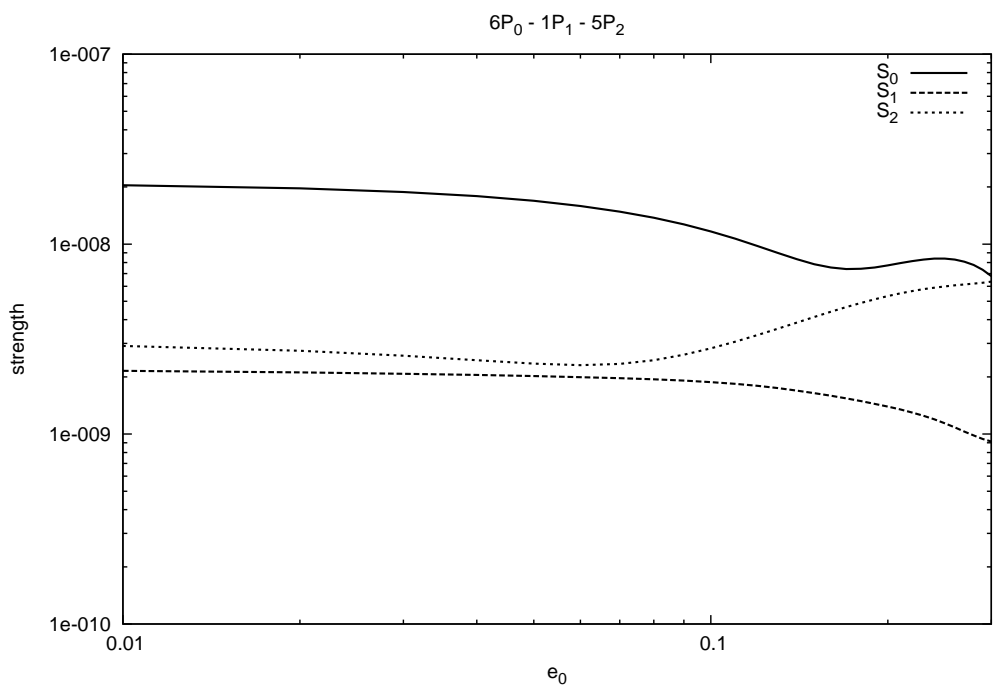}}
\caption{Strengths as function of $e_0$ for the zero order 3BR $6-1-5$. All other eccentricities are equal to 0.1 and all inclinations are zero.
For small eccentricities, the strength for each planet is almost independent of the eccentricity, as is typical for zero order resonances.}
\label{stexc0}
\end{figure}

\begin{figure}[h]
\resizebox{1.0\textwidth}{!}{\includegraphics{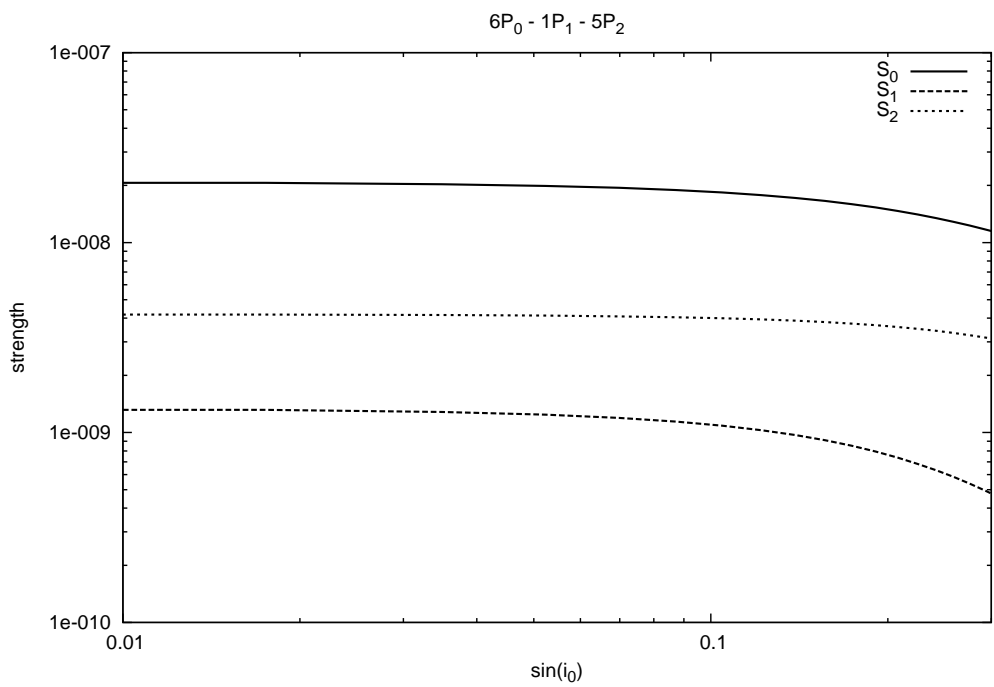}}
\caption{Strengths as function of $i_0$  for the same 3BR of Fig. \ref{stexc0}. All are circular orbits. All other inclinations are equal to $5.7^{\circ}$.
The strength for each planet is almost independent of the inclination as is typical for zero order resonances.}
\label{stinc0}
\end{figure}

Fig.  \ref{stexc0} shows the strengths as function of $e_0$ for the zero order resonance  $6-1-5$ located at  $2.2895$ au . They are almost independent of
the eccentricity in the range $0<e_0<0.1$.
Fig.  \ref{stinc0} shows the strengths as function of  $i_0$ and in analogy to the previous figure we
 can check they are almost independent of
the inclinations in the range $0<\sin (i_0) <0.1$. Zero order resonances exhibit this property of being almost independent of $e$ and $i$,
at least in the range of low eccentricity and low inclination orbits.

For zero order resonances we can compare our results with the theory by \citet{qu11} for closely spaced planetary systems. For example we calculated the strengths for the hypothetical system composed by $a_1=1$ au, $a_0=1.1586$ au and
$a_2=1.4$ au which $a_0$ corresponds to the resonance $2-1-1$ assuming coplanar and circular orbits. With our algorithm we obtained $S_2/S_1=1.3$ and $S_0/S_1=3.8$ while using formulae (29) by \citet{qu11} we obtain
$\Delta a_2 / \Delta a_1 = 1.2$ and $\Delta a_0 / \Delta a_1 = 2.1$, results in reasonable agreement taking into account the very different approaches involved and that in principle there is not necessarily a linear relation between strength
and $\Delta a$.

\subsection{Conclusions about the algorithm}
The algorithm presented in the Appendix provides reasonable estimates of the resonances strengths on each of the three massive bodies involved in a 3BR.
Also, the behaviour of the functions $S$ are coherent with known analytical results.
In particular, the behaviour of the strengths with the masses, orbital elements and order
of the resonances  is in agreement with that we can expect from the analytical expression of the disturbing function.
The examples presented above show that the resonance strength for a given planet is proportional to the masses of the two other planets as expected; thus
 the planet with the lowest mass is the one most affected by the resonance but for comparable masses the planet in the middle is in general the one that suffers the
 greatest dynamical effects.
 For low eccentricity and inclination orbits there is a neat dependence of the strengths with $e,i$ and the order.
 For a system with excited orbits in $(e,i)$, the dependence of the strengths with $(e,i)$ is not so clear mathematically and almost constant in the range (0,0.1) in $e$ and $\sin i$. For planetary or satellite systems with near circular near coplanar orbits only zero order resonances are dynamically relevant but for excited
systems, resonances of higher order may also be dynamically relevant.

\section{Numerical experiments}
\label{numeric}

In this section we explore the dynamical properties of some 3BRs by means of numerical methods and we compare the results with the predictions of our algorithm. The numerical integrations were carried out
with adaptations of the code EVORB \citep{fgb02}.

\subsection{Defining domains in a,e,i with dynamical maps}

\begin{figure}[h]
\resizebox{1.0\textwidth}{!}{\includegraphics{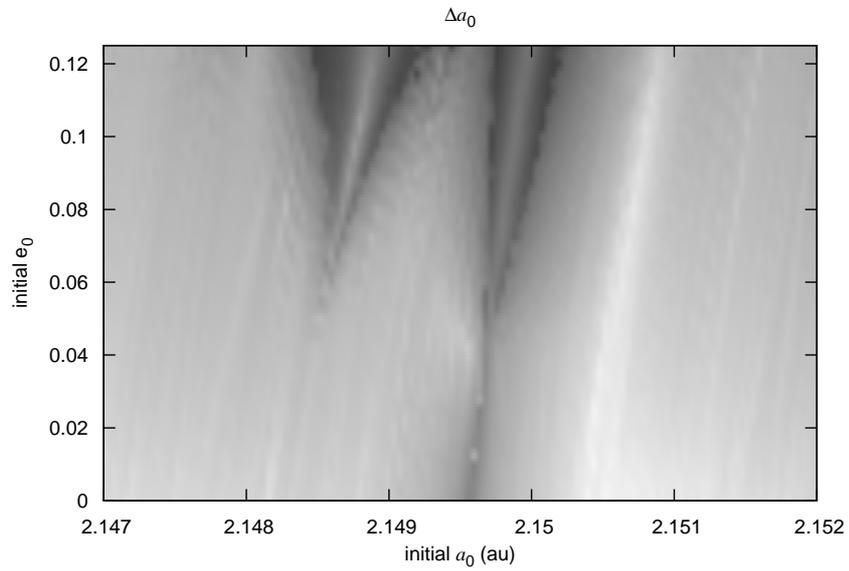}}
\caption{Dynamical map showing the domains of the resonances in $(a_0,e_0)$ for $e_1=e_2=0.01$, $i_i=1^{\circ}$ and $m_i=0.0001 M_{\odot}$. At the left is the 2BR $6P_0 - 13P_2$ and at right
the 3BR  $5- 1 -4$. Dark regions correspond to large amplitude ($10^{-4}$ au) oscillations of $\bar{a}_0$ and light regions to small amplitude ($10^{-7}$ au) oscillations. When varying $m_0$ this map remains unchanged. On the other hand, when increasing $m_1$
the 3BR increases its domain while the 2BR remains unchanged.}
\label{mapa}
\end{figure}

\begin{figure}[h]
\resizebox{1.0\textwidth}{!}{\includegraphics{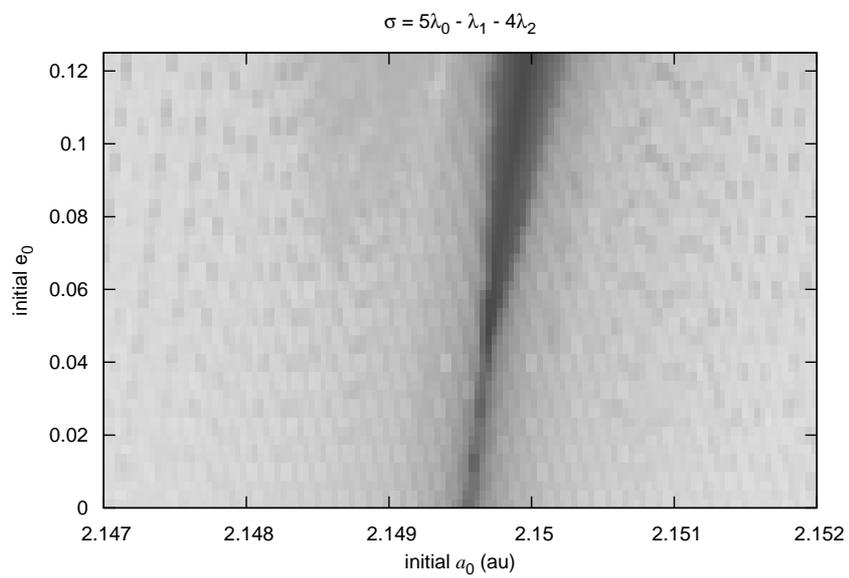}}
\caption{Behaviour of the critical angle $\sigma = 5\lambda_0 - \lambda_1 - 4\lambda_2$ corresponding to the 3BR for the same domain of Fig. \ref{mapa}. Black regions correspond to librations and they match very well with the domain of the 3BR showed in Fig. \ref{mapa}.}
\label{mapsigma3}
\end{figure}

\begin{figure}[h]
\resizebox{1.0\textwidth}{!}{\includegraphics{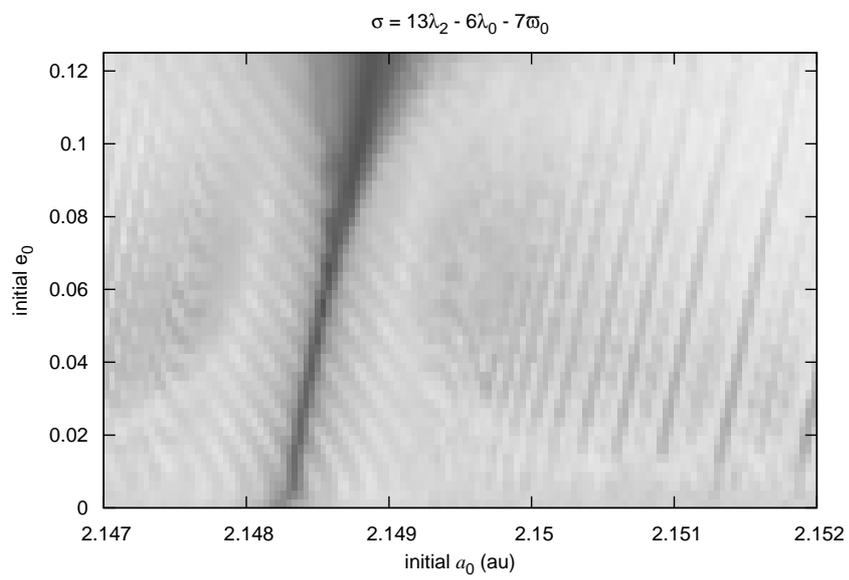}}
\caption{Behaviour of the critical angle $\sigma = 6\lambda_0 - 13\lambda_2 + 7\varpi_0$  corresponding to the 2BR for the same domain of Fig. \ref{mapa}. Black regions correspond to librations and they match very well with the domain of the 2BR showed in Fig. \ref{mapa}.}
\label{mapsigma2}
\end{figure}

We implemented codes in FORTRAN to construct dynamical maps near some 3BRs. In particular, from Fig.  \ref{atlas2y3} we choose the resonance $5-1-4$ near $a \simeq 2.15$ au. The dynamical maps in $(a,e)$ were constructed taking a grid of $\simeq 10000$ initial conditions for $P_0$ and calculating the time evolution of its semimajor axis  over a small number of libration periods, which it is about 15000 yrs. We calculate the mean $\bar{a}$ over a period of 1000 yrs in order to remove short period oscillations and moving this window over the entire integration we obtain $\bar{a}(t)$, which  approximately represents the time evolution of the semimajor axis due to the resonance's dynamical effects. Then, we calculate the amplitude $\Delta \bar{a} = \bar{a}_{max} - \bar{a}_{min}$ and plot this value as function of the initial $(a,e)$ with a gray scale from white to black according to
increasing values of $\Delta\bar{a}$. The structures that appear in Fig.  \ref{mapa} are due to the dynamical effects of
two resonances:
the one at the right is due to the 3BR $5-1-4$ and the one at the left is due to the high order 2BR $6P_0 - 13P_2$.
Each one shows a central region due to small amplitude oscillations, a dark border region with large amplitude oscillations near the separatrix and
an exterior region outside the resonance with near zero amplitude oscillations typical of a secular evolution. To identify these resonances we
implemented another code that calculates the corresponding critical angles during the same time interval of the numerical integration and performs a statistical analysis comparing the computed values of the critical angle with a uniform distribution between 0 and 360 degrees. Large departures from the uniform distribution, meaning small amplitude librations, are represented with black pixels and small departures, meaning large amplitude or circulations,
are represented with white pixels. The resulting map for the
critical angle $\sigma = 5\lambda_0 - \lambda_1 - 4\lambda_2$ is showed in Fig.  \ref{mapsigma3} and the map for
$\sigma = 6\lambda_0 - 13\lambda_2 + 7\varpi_0$ is showed in Fig.  \ref{mapsigma2} which confirm that the dynamical effects showed in Fig.  \ref{mapa} are
due to these resonances. It is interesting to note that at small eccentricities both critical angles librate but, looking at Fig.  \ref{mapa},
we can verify that the high order 2BR have null dynamical effects
meanwhile the zero order 3BR have appreciable effects in semimajor axis even at zero eccentricities, which is in agreement with the theory and the predictions of our semianalytical  method. At near zero eccentricities  the dynamically relevant resonances are only those of order zero.

\begin{figure}[h]
\resizebox{1.0\textwidth}{!}{\includegraphics{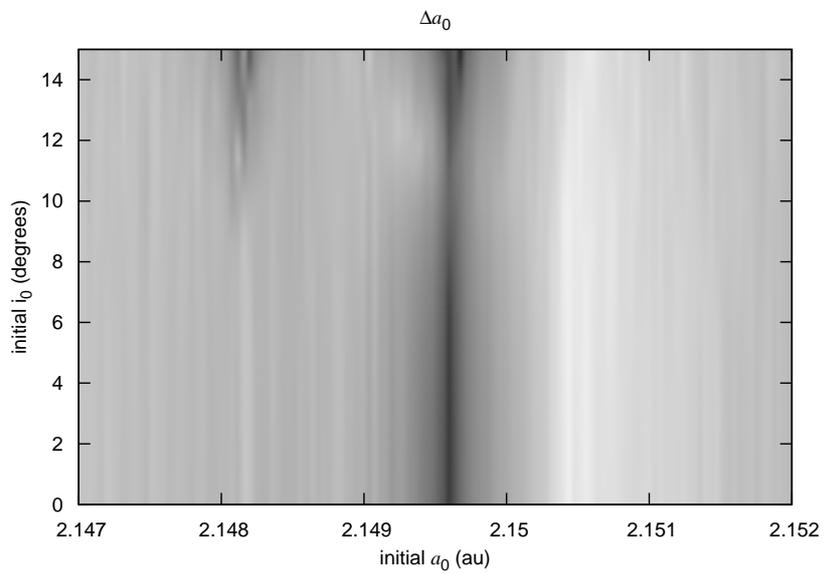}}
\caption{Dynamical map showing the domains of the resonances in $(a_0,i_0)$ for $e_i=0.01$, $i_1=i_2=0.1^{\circ}$.
Dark regions correspond to large amplitude ($10^{-5}$ au) oscillations of $\bar{a}_0$ and light regions to small amplitude ($10^{-7}$ au) oscillations. The domain of the 3BR is unaffected by $i_0$ while the 2BR shows up only for large inclinations.}
\label{mapi}
\end{figure}

The resonance domain in $(a,i)$ is represented in the map given in Fig.  \ref{mapi} which was calculated taking $e_i=0.01$ and $i_1=i_2=0.1^{\circ}$. Both
 resonances can be distinguished and as we have remarked previously the domain of the zero order 3BR is almost independent of the inclination while the high order 2BR is strongly dependent on inclination being vanishingly small at low inclinations.

\begin{figure}[h]
\resizebox{1.0\textwidth}{!}{\includegraphics{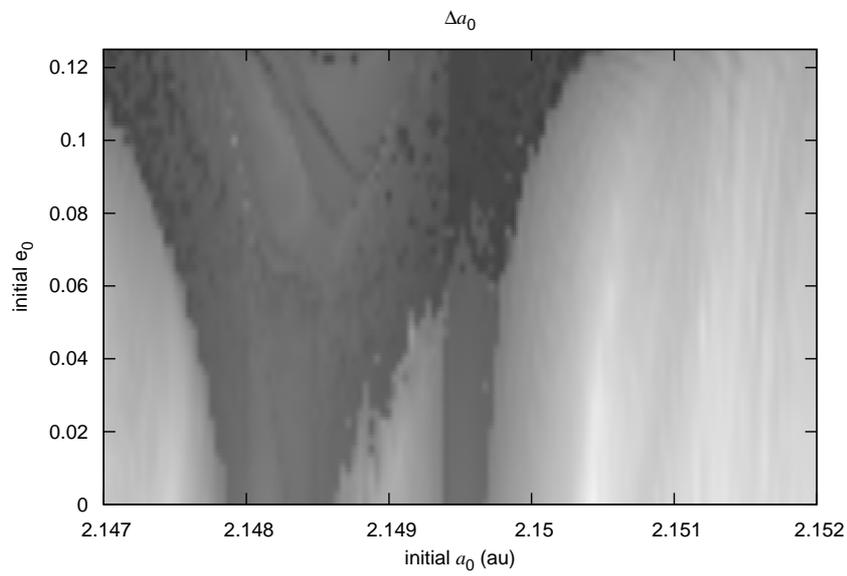}}
\caption{Same as Fig. \ref{mapa}
but for an excited system with $e_1=e_2=0.1$, $i_i=10^{\circ}$.
Dark regions correspond to large amplitude ($10^{-3}$ au) oscillations of $\bar{a}_0$ and light regions to small amplitude ($10^{-7}$ au) oscillations.
The 2BR has grown by a very large amount erasing the traces of the 3BR for $e_0 > 0.06$.}
\label{mapexc}
\end{figure}

The last map presented in Fig.  \ref{mapexc} was generated for a dynamically excited system and shows an impressive growth of the 2BR
that overrides the 3BR for $e_0>0.05$ illustrating that for excited systems 2BRs must dominate over the 3BRs. Note also that the domain of the 3BR is almost independent of the eccentricity. Nevertheless,
back to Fig. \ref{mapa}, it is supposed that a zero order 3BR must be almost independent of the eccentricity while it is showed some growing of its domain for $e_0 > 0.06$ not only in Fig. \ref{mapa} but also in Fig. \ref{mapsigma3}. This seems contradictory with our results  for zero order 3BRs. We have checked that there are not 2BRs superposed to the 3BR, then we can conjecture that the multiplet of this resonance generate that feature. The multiplet is composed by a principal term independent of $e$ plus several terms depending on powers of the eccentricity.

\subsection{Libration properties and dynamical evolution in a migrating scenario}

\citet{li11} studied the capture of a system in a chain of 2BRs due to a migration scenario. For the initial conditions they considered, they found that, as a general rule,
the two inner planets are captured in a 1:2 resonance and the third planet is captured in the 1:2 or 1:3 resonance with the middle one. This configuration
allows a very interesting evolution in eccentricity and inclination and the resulting 3BR is in fact a superposition of 2BRs.
In particular the semimajor axes evolve as expected for two planets locked in a 2BR maintaining a constant ratio as they evolve towards the star due to the migration mechanism. In this paper we are interested in detecting dynamical mechanisms generated by pure 3BRs, that means not reducible to a superposition of 2BRs which, in general, are stronger and then they could erase the effects of the 3BRs.

\begin{figure}[h]
\resizebox{1.0\textwidth}{!}{\includegraphics{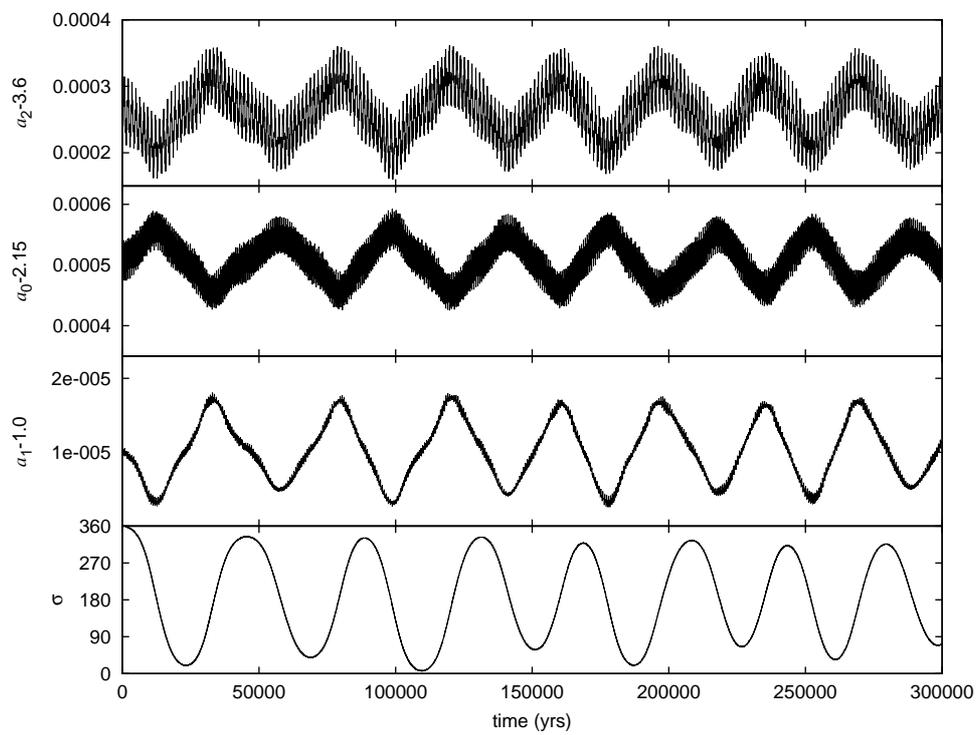}}
\caption{Time evolution of a planetary system inside the zero order pure 3BR $5 - 1 - 4$ with
$P_1$ and $P_2$ at the same positions of the previous figures.
Initial values are $e_i=0.01$ and $i_i=1^{\circ}$. The mean semimajor axes were calculated with a moving window of 500 years. The critical angle is  $\sigma = 5\lambda_0 - \lambda_1 - 4\lambda_2$. The oscillations of the planet in the middle are opposed to the oscillations of the other two planets.}
\label{mig1}
\end{figure}

A dynamical evolution of a pure 3BR is exemplified  in Fig.  \ref{mig1} where we show the time evolution of the mean $a_i$ together with the time evolution of the critical angle
 $\sigma = 5\lambda_0 - \lambda_1 - 4\lambda_2$ for the same working planetary system we have idealized in Fig. \ref{atlas2y3} with initial conditions near the border of this zero order 3BR.
 Mean $a_i$ were calculated with a running window of 500 years.
 The time evolution of the semimajor axes are in agreement with the theoretical results by \citet{qu11} who showed that,  at least for zero order 3BRs, the exterior and interior planet
 have semimajor axes oscillating in phase and
 the planet in the middle is half period shifted. Also, running our algorithm for this case we obtain
$S_0/S_1 \simeq 13$ and $S_0/S_2 \simeq 4$ which can be compared with the $\Delta a$ taken from Fig. \ref{mig1}:
$\Delta a_{0}/\Delta a_{1}\sim 10$ and $\Delta a_{0}/\Delta a_{2}\sim 1$.
It is not an exact match because  $S_i$ probably is not directly proportional to $\Delta a_{i}$, but we can conclude our strengths $S_i$ are coherent with the dynamical effects observed in the semimajor axes of the involved bodies.

\begin{figure}[h]
\resizebox{1.0\textwidth}{!}{\includegraphics{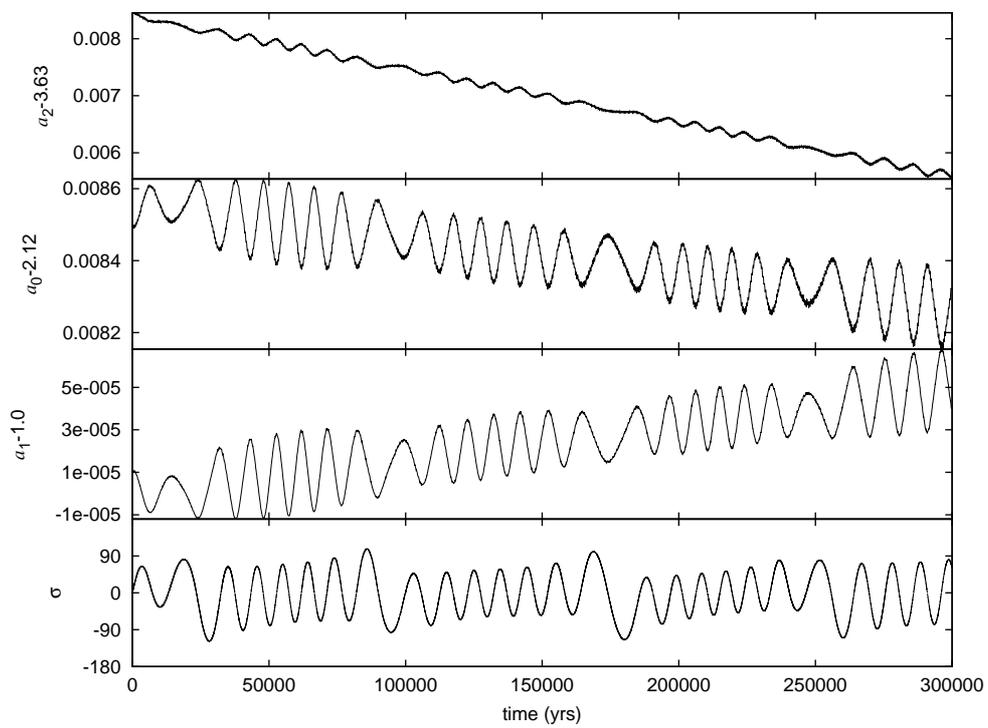}}
\caption{Time evolution of a planetary system inside the pure first order 3BR $4 - 1 - 2$ imposing a forced inward
migration for $P_2$. Initial values are $e_i=0.01$ and $i_i=1^{\circ}$. Mean semimajor axes calculated with a moving window of 500 years.
 The critical angle is $\sigma = 4\lambda_0 - \lambda_1 - 2\lambda_2 - \varpi_1$. A system inside a 3BR when forced to migrate can exhibit rate changes positive and negative for the semimajor axes.}
\label{mig2}
\end{figure}

 In the next numerical experiment we simulate a migration of the exterior planet $P_2$ towards the central star while the system evolves inside the first order 3BR
 $4 - 1 -2$. The migration of the outer planet $P_2$ is imposed by an artificial constant force with direction contrary to the orbital velocity generating a variation rate of  $\dot{a} = -1\times 10^{-8}$ au/yr. The resulting evolution is showed in Fig.  \ref{mig2} where undoubtedly the three semimajor axes evolve in synchrony with the oscillations of the critical angle and is again verified that the oscillations of the semimajor axes of the exterior and interior planets are in phase while the planet in the middle is shifted half a period.
 Also, a not very intuitive phenomena  is observed:  while the outer two planets  migrate inwards the inner planet
 $P_1$ migrates outwards. Contrary to the case of systems captured in a 2BR where, in general, both semimajor axes must grow or decrease simultaneously linked by the resonant relation, in the case of 3BRs there is another degree of freedom that allows this behaviour.
 Nevertheless, it is important to mention that for planetary systems with very low eccentricity orbits trapped in 2BRs it is possible to observe divergence of orbits as showed by \citet{bm13}
 because for low eccentricity orbits the location of the resonance is shifted in semimajor axes due to the Law of Structure of the resonance \citep{sfm88}
 which is a dependence of $a_{res}$ on the orbital eccentricity. Then, if the eccentricities change, being small, the ratio $a_1/a_2$ of a pair of planets locked in resonance can change due to the Law of Structure.
We have simulated other migrations processes with systems inside other pure 3BRs and we have obtained that is very common that the planets migrate with diverging orbits
not only for low eccentricity orbits but also for excited orbits.

\begin{figure}[h]
\resizebox{1.0\textwidth}{!}{\includegraphics{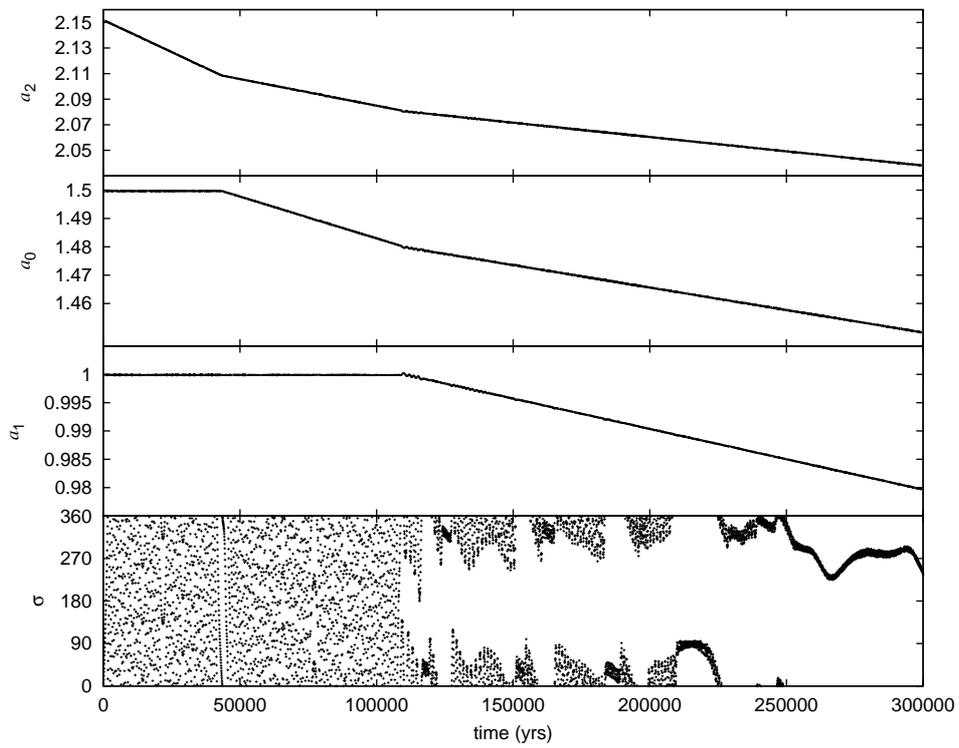}}
\caption{Capture and evolution in a chain of two 2BRs imposing a forced inwards migration of $P_2$. The critical angle showed at bottom is $\sigma = 3\lambda_0 - \lambda_1 -2\lambda_2$. Mean semimajor axes calculated with a moving window of 1000 years. A system captured in a chain of 2BRs when forced to migrate in general exhibit rate changes of the same sign for the semimajor axes.  }
\label{2mas2}
\end{figure}

We performed a series of numerical experiments trying to capture a planetary system in a 3BR from outside the resonance in a migrating scenario using migration rates from $10^{-9}$ to  $10^{-6}$ au/yr both positive and negative.
Our very preliminary results indicate that capture in a pure 3BR is a very rare event while capture in a chain of 2BRs is a very frequent result
as has been demonstrated by \citet{li11}. An example of this last case is showed
in Fig. \ref{2mas2} where an external migrating planet $P_2$ with $\dot{a} = -1\times 10^{-6}$ au/yr captures the middle planet $P_0$ in the
$3P_0 - 5P_2$ resonance at $t=42000$ yrs and then $P_0$ captures the planet $P_1$ in the resonance $9P_0 - 5P_1$ at $t=111000$ yrs. Consequently
the system gets trapped in the zero order 3BR $3-1-2$ which is the lowest order 3BR that can be obtained from the two 2BRs, but its dynamics is mostly due to the superposition of the mutual 2BRs.
In this example all three planets have $m=10 M_{\oplus}$, initial $e_i=0.01$ and mutual inclinations of
about 1 degree. Our experiments show systematically that when the system cross a 3BR the three planets experience a jump in semimajor axes: the planets in the extremes have a jump in the same direction and the planet in the middle in the contrary direction, in agreement with \citet{qu11}. The capture in pure 3BRs deserves much more study and is beyond the scope of this work.

\subsection{Application to the Galilean satellites}
\label{sectiongal}

\begin{figure}[h]
\resizebox{1.0\textwidth}{!}{\includegraphics{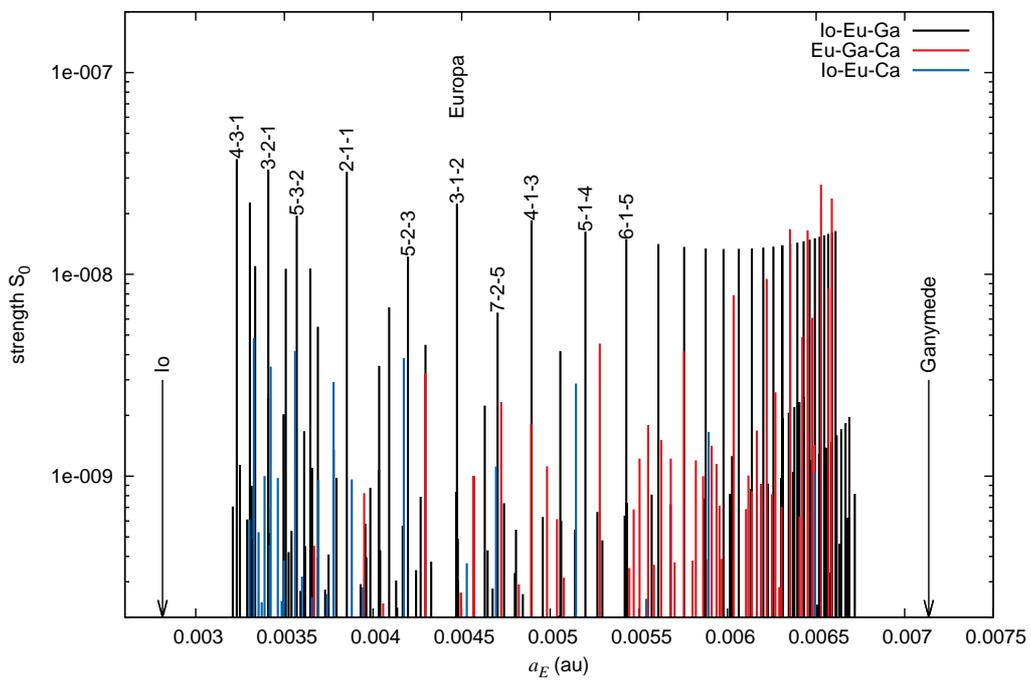}}
\caption{Atlas of 3BRs between Io and Ganymede. Three body resonances involving Io-\textit{Europa}-Ganymede in black,  \textit{Europa}-Ganymede-Callisto in red and  Io-\textit{Europa}-Callisto in blue where \textit{Europa} is an hypothetical object with the same mass and orbital elements of the actual Europa but with a semimajor axis defined by the resonance.  Some resonances involving Io-\textit{Europa}-Ganymede are labeled. Positions of Io and Ganymede are indicated with arrows. }
\label{galilean}
\end{figure}

We applied our semi-analytical method to explore all possible 3BRs between Galilean satellites near the location of Europa.
The method assumes there are no 2BRs between them, which it is not the case because of the existence of the resonance
2:1 between Io and Europa and also 2:1 between Europa and Ganymede. Then, the results we obtained for the 3BRs involving Io, Europa and Ganymede must be taken with caution.
We started taking the two fixed bodies Io and Ganymede as $P_1$ and $P_2$ respectively, with its present orbits and masses taken from Table \ref{table1} and we calculated
all relevant 3BRs located in between both satellites as experienced by a third body, $P_0$, with the same mass and orbital parameters of Europa, except
for its semimajor axis which is defined by the different resonances we are trying to evaluate. The resulting set of resonances with their strengths is showed in Fig \ref{galilean} with black lines.
As we expected, the actual Europa is located in the resonance $3E-1I -2G$, or $3-1-2$ in our notation, which is one of the strongest 3BRs of the system. Then, we repeated the method but considering Ganymede as $P_1$ and Callisto
as $P_2$ and we calculate all relevant 3BRs involving these satellites with an hypothetical Europa. Finally, we repeat the procedure but taking
Io as $P_1$ and Callisto as $P_2$. All three sets of 3BRs are plotted in Fig.  \ref{galilean}.   The resonance  $3E-1I -2G$
is one of the strongest resonances located in a region relatively  devoid of other perturbing 3BRs. It is possible to distinguish in the figure the second order 3BR $1E-1I+2G$ almost superimposed with $3E-1I -2G$
but with strength $S<1\times 10^{-9}$.

If we look at the ratios of the strengths that our method predicts for the resonance
  $3E-1I -2G$ we find $S_{E}/S_{I}\sim 6$ and  $S_{E}/S_{G}\sim 12$ which can be compared with the ratios between the $\Delta a$ obtained from the numerical integrations given, for example, in \citet{mu02} which are $\Delta a_{E}/\Delta a_{I}\sim 4$ and  $\Delta a_{E}/\Delta a_{G}\sim 6$.
Note that in \citet{mu02} $a_{I}$ and  $a_{G}$  are evolving in phase while for the  body in between, $a_{E}$,  is shifted by half a libration period in agreement with
\citet{qu11}.

\begin{figure}[h]
\resizebox{1.0\textwidth}{!}{\includegraphics{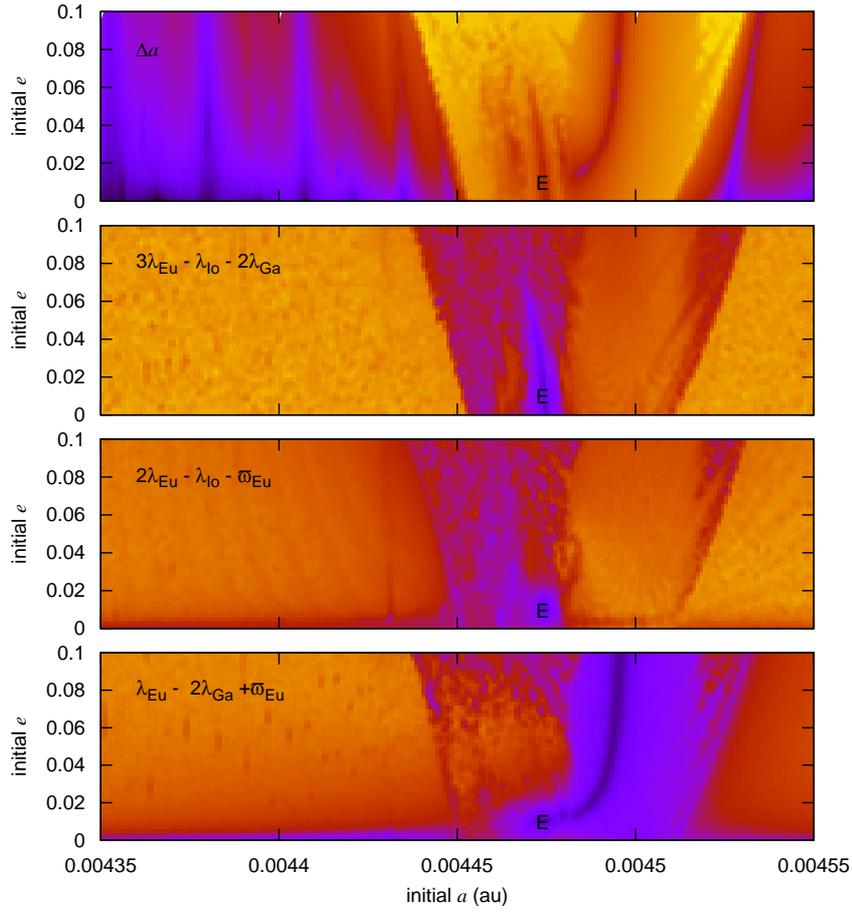}}
\caption{Dynamical maps near the location of Europa, indicated with E.  $\Delta \bar{a}$ in top panel
is showed with
dark regions corresponding to small amplitude ($10^{-9}$ au) oscillations of $a_0$ and with vivid color regions corresponding to large amplitude ($10^{-5}$ au) oscillations.
The critical
angles $3\lambda_{E}-\lambda_{I} -2\lambda_{G}$ in second panel, $2\lambda_{E}-\lambda_{I} -\varpi_{E}$ in third panel
and $\lambda_{E}-2\lambda_{G} +\varpi_{E}$ in bottom panel. Dark colors correspond to small amplitude oscillations of the critical angles and vivid colors to large amplitude oscillations and circulations.}
\label{europamap}
\end{figure}

In order to distinguish the different resonances affecting Europa we have constructed dynamical maps for a simple model consisting of a system of the four satellites orbiting Jupiter with its orbital elements taken from Table \ref{table1} and
considering Jupiter's oblateness through the $J_2$ term. The map for $\Delta \bar{a}$
in Fig. \ref{europamap} top panel
was constructed by means of numerical integrations for intervals of 15 years and using a moving window of 1 year to calculate $\bar{a}(t)$. The map is constructed with 10000 different initial conditions taken from a grid in $(a,e)$ and we compared this map with the map obtained from the evolution of various critical angles. The dynamical map of Fig. \ref{europamap} top panel
shows with vivid colors large $\Delta \bar{a}$
associated with the borders of the resonance region
and with dark colors small values of $\Delta \bar{a}$
associated with small amplitude librations in the center of the resonance or with no oscillations, typical of secular non resonant evolution
outside the resonance.
The maps for the critical angles show regions of small amplitude librations with dark colors and circulations with vivid colors.
There is a close correlation between the dynamical map for $\Delta \bar{a}$ with the evolution of the critical angles
$3\lambda_{E}-\lambda_{I} -2\lambda_{G}$ of the 3BR, $2\lambda_{E}-\lambda_{I} -\varpi_{E}$ of the exterior 2BR
$2E-1I$ and
 $\lambda_{E}-2\lambda_{G} +\varpi_{E}$ of the interior 2BR $1E-2G$.

From examination of Fig. \ref{europamap} we can conclude that the features in the map for $\Delta \bar{a}$ in the left region,
between $a = 0.00444$ au and $a = 0.00448$ au, have a very good match with the features in the map of
the 3BR in Fig. \ref{europamap} second panel. This map shows that Europa is located in the very central region of the 3BR, region which has low dependence with
the eccentricity as is typical for a zero order 3BR.
The zone at the right of $a = 0.00448$ au  matches with the features in the map for
the 2BR 2:1 between Europa and Ganymede at bottom panel in Fig. \ref{europamap}.
In this panel it is possible to identify the Law of Structure of the resonance 2:1, which is the deviation of the location of the exact resonance from the nominal value $a_{res}$ at low eccentricities as we have explained above.
For completeness we show in the third panel the map for the corresponding critical
angle for the exterior 2BR 1:2 between Io and Europa which seems to have no relevant effects
in the map for $\Delta \bar{a}$.
 In our numerical integrations a particle with the same orbital elements of Europa shows librations in the three critical angles but the largest oscillations in $a_{E}$ are correlated with the critical angle of the 3BR $3E-1I -2G$. Then, Fig. \ref{europamap}
suggests that Europa is mostly dominated by the pure 3BR.

\begin{figure}[h]
\resizebox{1.0\textwidth}{!}{\includegraphics{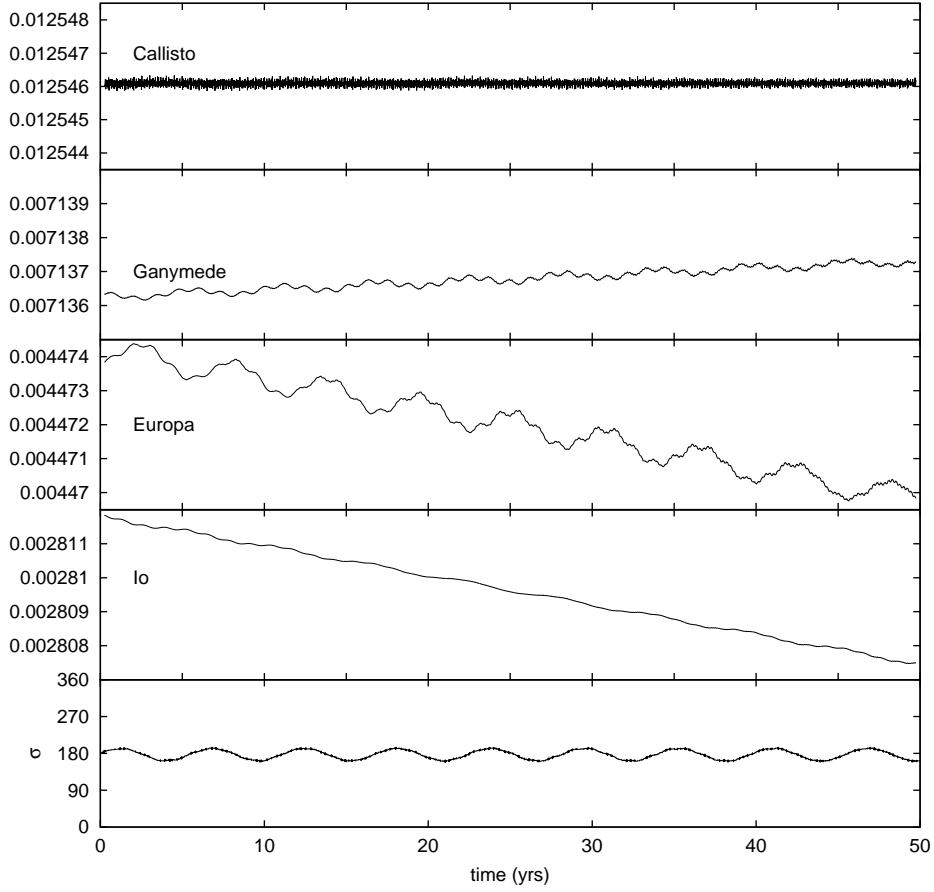}}
\caption{Mean semimajor axes of Callisto, Ganymede, Europa and Io expressed in au evolving due to an induced arbitrary inwards migration of Io. Mean semimajor axes calculated using running window of 0.5 years. In bottom panel the corresponding evolution of the critical angle of the Laplacian resonance. The short period small amplitude oscillations in semimajor axes are correlated with the time evolution of the critical angle of the resonance $1E - 2G$.}
\label{galmigration}
\end{figure}

Various attempts have been done in order to identify possible migrations of the Galilean satellites due to tidal effects caused by Jupiter and, consequently, to understand the future of the Laplacian resonance \citep{la09}.
Fitting the parameters of a very complete physical model to a large set of astrometric positions \citet{la09} conclude that,
due to the mechanism of tides in Jupiter-Io system, at present Io is migrating
inwards to Jupiter at a very low rate ($\dot{a} \simeq -2.6\times 10^{-14}$ au / yr) while Europa and Ganymede migrate outwards, being the system leaving the exact commensurability of the Laplacian resonance.
In order to evaluate the strength of the Laplacian resonance and, in particular, if it is capable of surviving to a migration mechanism we performed a numerical simulation of the system given by Jupiter plus the four Galilean satellites with an
imposed inwards migration for Io
given by  $\dot{a} = -1\times 10^{-7}$ au / yr, that means approximately seven orders of magnitude greater than the deduced by \citet{la09}.
If the 3BR is not strong enough it will be broken by the imposed artificial migration, otherwise Europa and Ganymede will migrate in such a way that the
resonant relation is conserved.
The resulting evolution of the system is given in Fig.  \ref{galmigration}.
In our simulation Europa responds migrating inwards like Io but Ganymede moves outwards; while Callisto does not experience orbital changes as is expected
because it is not participating in the Laplacian resonant mechanism.
All critical angles remain librating during the integration period but while the libration amplitude of the two 2BRs
increase with time, the amplitude of the Laplacian resonance remains constant,  see Fig. \ref{galmigration} bottom panel.
Differences with results by \citet{la09} can be explained because the models are very different, but it is remarkable that in our experiment the 3BR persists.
The largest amplitude oscillations in the three semimajor axes we see in Fig. \ref{galmigration}  are linked to
the librations of the Laplacian 3BR and the high frequency low amplitude oscillations are linked to the librations of $\lambda_{E}-2\lambda_{G} +\varpi_{E}$, suggesting that the main dynamical mechanism is the 3BR and that
the resonance $1E-2G$ only makes a small contribution. This is consistent with the information we can deduce from Fig. \ref{europamap}.
Our Io-migrating experiment does not pretend to show the actual dynamical evolution of the satellite system, just try to demonstrate that, even
in case the 2BRs were breaking, the Laplacian 3BR is strong enough to survive, even imposing migration rates several order of magnitude greater than actual ones.

\begin{table}
  \centering
 \begin{tabular}{  l   r    r  r  r r r r }
  \hline
 satellite  &    $a$ (au) &   $e$       &   $i (^{\circ})$     & $\Omega (^{\circ})$    & $\omega (^{\circ})$  &    $M (^{\circ})$         &   $m$ $(M_{\odot})$ \\
 \hline
   Io       & 0.002812  &   0.0041  &  0.036  &    43.977  &    84.129 &  342.021      &   4.5D-08 \\
  Europa    & 0.004474  &   0.0094  &  0.466  &   219.106  &    88.970 &  171.016      &   2.4D-08 \\
  Ganymede  & 0.007136  &   0.0013  &  0.177  &    63.552  &   192.417 &  317.540      &   7.6D-08 \\
  Callisto  & 0.012551  &   0.0074  &  0.192  &   298.848  &    52.643 &  181.408      &   5.4D-08 \\
  \hline	 								
\end{tabular}	 								
 \caption{Mean orbital elements for Epoch 1997 Jan. 16.00 TT taken from ssd.jpl.nasa.gov. $J_2=14.7\times 10^{-3}$ }\label{table1}			
\end{table}

\section{Concluding remarks}
\label{conclu}

Three body resonances between massive bodies generate different effects on each planet.
The semianalytical method presented here have proven useful to predict the configurations and approximate strengths of the 3BRs generated by a system of three massive bodies with arbitrary orbits which are not in 2BRs between them and provides a useful tool for evaluating the dynamical relevance of 3BRs among planetary and satellite systems.
It allows to analyze the dependence of the strengths  on each planet with the masses, eccentricities and inclinations of
 all involved planets. In particular, the dependence with the inclinations can now be explored for the first time.
 For near zero eccentricity orbits, zero order 3BRs are the strongest ones, even stronger than 2BRs in some cases.
On the other hand, for excited systems, first or second order 3BRs could be equally relevant than zero order 3BRs, but if 2BRs
are present in the neighborhood they will dominate.
  When we compared the effect on each planet participating in a resonance, the most affected one is that of smallest mass or, in general, the planet in the middle when the masses are similar. We confirmed that the planet in the middle has oscillations in $a$ shifted half a libration period with respect to the other planets.
 It is common that in a migrating scenario one of the bodies locked in a pure 3BR migrate in the opposite direction than the other two due to the existence of a new degree of freedom in the equation linking the mean motions.
Our very preliminary results of our numerical simulations suggest that capture in a pure 3BR is an unusual event but, on the other hand, systems captured in pure 3BRs can survive to imposed migration mechanisms.
Our study of the case of the Galilean satellites suggest that the Laplacian 3BR dominate the system and is strong enough to maintain the system locked
 in resonance even for migration rates several orders of magnitude greater than the deduced by present theories.

\bigskip

\textbf{Acknowledgments.}
We acknowledge support from PEDECIBA and Project CSIC Grupo I+D 831725 - Planetary Sciences.
We thank the reviewers that contributed to clarify various points of the original manuscript.


\clearpage

\appendix
\section{Numerical approximation to the disturbing function for planetary three body resonances}

Given two planets $P_1$ and $P_2$ in arbitrary orbits, the mean resonant disturbing function, $\mathcal{R}(\sigma)$, that drives the resonant motion of the planet $P_0$
assumed inside an arbitrary 3BR
could be ideally calculated eliminating the short period terms of the
resonant disturbing function $R$ for the planet by means of
\begin{equation}\label{doble}
  \mathcal{R}(\sigma) =  \frac{1}{4\pi^2}\int_{0}^{2\pi}d\lambda_1\int_{0}^{2\pi}
    R\Bigl(\lambda_0(\sigma,\lambda_1,\lambda_2,\gamma),\lambda_1,\lambda_2\Bigr)  d\lambda_2
\end{equation}
where $\lambda_0$ was explicitly written in terms of the variables $\lambda_1,\lambda_2$ and the parameters $\sigma,\gamma$ using Eq. (\ref{sigmaj})
and
where $R(\lambda_0,\lambda_1,\lambda_2) = R_{01} + R_{02}$
being
\begin{equation}
\label{Rij}
    R_{ij}= k^{2} m_{j} ( \frac{1}{r_{ij}} - \frac{\vec{r_i} \cdot \vec{r_j}}{r_j^3} )
\end{equation}
where $k$ is the Gaussian constant, $m_j$ the mass of planet $P_j$ and $\vec{r_i}, \vec{r_j}$ are the astrocentric  positions of bodies with subindex $i$ and $j$ respectively.
Note that for each set of values $(\sigma,\lambda_1,\lambda_2,\gamma)$ there are $k_0$ values of $\lambda_0$ that satisfy Eq. (\ref{sigmaj}),
which are:
\begin{equation}\label{lambdas2}
    \lambda_0 =  \left(\sigma - k_1\lambda_1 -k_2\lambda_2 - \gamma \right)/k_0 + n 2 \pi / k_0
\end{equation}
with $n=0,1,...,k_0-1$. All them contribute to $\mathcal{R}(\sigma)$ in Eq. (\ref{doble}) so we have to evaluate all these $k_0$ terms and calculate the mean,
which is equivalent to integrate in $\lambda_0$ maintaining the condition (\ref{sigmaj}).

The disturbing function of a 3BR is a second order function of the planetary masses,
which means  the calculation of the double integral (\ref{doble}) cannot be
done over the perturbing function evaluated at the unperturbed astrocentric positions.
To properly evaluate the integral it is necessary to
take into account their mutual perturbations in the position vectors   $\vec{r_i}$.
Two body mean motion resonances are a simpler case because being a first order
perturbation in the planetary masses the position vectors
can be substituted by the Keplerian, non perturbed positions.

In order to estimate the behavior of $\mathcal{R}(\sigma)$,
\citet{ga14} adopted the following scheme for computing the double integral of Eq. (\ref{doble}):
\begin{equation}
 R(\lambda_0,\lambda_1,\lambda_2)  \simeq     R_u + \Delta R
\end{equation}
where $R_u$ stands from $R$ calculated at the unperturbed positions of the three bodies
and $\Delta R$ stands from the variation in $R_u$ generated by the perturbed (not Keplerian) displacements of the three bodies
in a  small interval $\Delta t$.
More clearly, given any set of the three position vectors $\vec{r_i}$ satisfying Eq. (\ref{sigmaj}) we compute the mutual perturbations
of the three bodies
and calculate the $\Delta\vec{r_i}$ that they generate in a small interval $\Delta t$ and the $\Delta R$ associated.
This scheme is equivalent to evaluate the integral over the infinitesimal trajectory
the system follows due to the mutual perturbations when released at all possible unperturbed positions that verify Eq. (\ref{sigmaj}).
We have then
\begin{eqnarray}
\label{R}
    R_u &=&   R_{01} +   R_{02} \\
    \Delta R &=& \Delta R_{01} + \Delta R_{02}
\end{eqnarray}
where $R_{01}$ and $R_{02}$ refer to the disturbing functions evaluated at the unperturbed positions and
$\Delta R_{01}$ and $\Delta R_{02}$ refer to the variations due to displacements caused by the mutual perturbations:

\begin{equation}
\label{dr01}
     \Delta R_{01} = \nabla_0 R_{01} \Delta \vec{r_0} +  \nabla_1 R_{01} \Delta \vec{r_1}
\end{equation}

\begin{equation}
\label{dr02}
     \Delta R_{02} = \nabla_0 R_{02} \Delta \vec{r_0} +  \nabla_2 R_{02} \Delta \vec{r_2}
\end{equation}
where $\Delta \vec{r_i}$  refers to displacements with respect to the astrocentric Keplerian motion
and being
\begin{equation}
\label{gradiRij}
    \nabla_i R_{ij}= k^{2} m_{j} (\frac{ \vec{r_j} -  \vec{r_i} }{r_{ij}^3}  - \frac{\vec{r_j}}{r_j^3})
\end{equation}
\begin{equation}
\label{gradjRij}
    \nabla_j R_{ij}= k^{2} m_{j} (\frac{ \vec{r_i} -  \vec{r_j} }{r_{ij}^3}  - \frac{\vec{r_i}}{r_j^3} + 3(\vec{r_i}\vec{r_j})\frac{\vec{r_j}}{r_j^5})
\end{equation}
From the equations of motion we have:
\begin{equation}\label{dx0dt2}
    \ddot{\vec{\Delta r_0}} =  \nabla_0 R_{01} +  \nabla_0 R_{02}
\end{equation}
\begin{equation}\label{dx1dt2}
    \ddot{\vec{\Delta r_1}} =  \nabla_1 R_{12} +  \nabla_1 R_{10}
\end{equation}
\begin{equation}\label{dx2dt2}
    \ddot{\vec{\Delta r_2}} =  \nabla_2 R_{21} +  \nabla_2 R_{20}
\end{equation}
Integrating twice we obtain the displacements with respect to the Keplerian motion:
\begin{equation}\label{dx0}
    \vec{\Delta r_0} \simeq  (\nabla_0 R_{01} +  \nabla_0 R_{02}) \frac{(\Delta t)^2}{2}
\end{equation}
\begin{equation}\label{dx1}
    \vec{\Delta r_1}  \simeq   (\nabla_1 R_{12} +  \nabla_1 R_{10}) \frac{(\Delta t)^2}{2}
\end{equation}
\begin{equation}\label{dx2}
    \vec{\Delta r_2} \simeq   (\nabla_2 R_{21} +  \nabla_2 R_{20}) \frac{(\Delta t)^2}{2}
\end{equation}
As the integral of $R_{u} = R_{01} + R_{02}$ becomes independent of $\sigma$,
we are only interested
in computing the function $\rho(\sigma)$ defined by
\begin{equation}\label{suma}
\rho(\sigma) = \frac{1}{4\pi^2}\int_{0}^{2\pi}d\lambda_1\int_{0}^{2\pi}\Delta R d\lambda_2
\end{equation}
always satisfying Eq. (\ref{sigmaj}). Its dimensions are $[M]^{2}k^{2}/[L]$ in solar masses, au and days.

Note that $\rho(\sigma)$ is a summation of terms each one factorized by two masses while in the case of 2BRs the disturbing function is proportional to only one planetary mass,
making 3BRs much weaker than 2BRs.
Note also that $\Delta R$ is calculated via some arbitrary  $\Delta t$ that we identify with the permanence time in each element of the phase space $(\Delta \lambda_0, \Delta \lambda_1, \Delta \lambda_2)$.
If the double integral is computed dividing the dominium in $N$ equal steps in $\lambda_1$ and
$N$ equal steps in $\lambda_2$ we can calculate the mean elapsed time  $\Delta t$ in the element of phase space
as
\begin{equation}\label{dt}
    \Delta t = \frac{\sqrt[3]{T_0 T_1 T_2}}{N}
\end{equation}
where $T_i$ are the orbital periods. Another way of understanding the meaning of $\Delta t$ is to calculate the probability of finding the system
in a particular configuration during  $\Delta t$, which is $(\Delta t)^{3}/(T_0 T_1 T_2)$.  Then
\begin{equation}\label{dtcua}
    \Delta t ^{2} = \frac{4\pi^{2} a_0 a_1 a_2}{k^2 M N^2}
\end{equation}
where $M$ is the mass of the central body (star or planet) expressed in solar masses.
Note that $N$ is an arbitrary integer but it must be always the same if we want to compare functions $\rho(\sigma)$ for different resonances.
 Taking $N$ equal for all resonances its
actual value is irrelevant; in our codes we use  $N=1$.
Considering $\sigma$ as a constant parameter we calculate the integral (\ref{suma}) for a set of values of $\sigma$ between $(0,2\pi)$
and we obtain numerically $\rho(\sigma)$.

We consider the strength of the resonance, $S$, the value  of the semiamplitude $S=\Delta \rho /2$ as in \citet{ga14}.
The reason for this definition is that if $\sigma$ generates large variations in $\rho$ is because it has some dynamical relevance. On the other hand, if variations in $\rho$ are
negligible is because the critical angle, that means the resonance, is irrelevant for the dynamics.

An important difference with the restricted case is that in the
general 3BR problem all three planets feel the resonance, then there are dynamical effects in all three planets. We calculate these
resonant effects in the other two planets following an analogue procedure than the one we followed for planet $P_0$.
While equations (\ref{dx0}) to (\ref{dx2}) are the same the corresponding $\Delta R$ are for planet $P_1$ :
\begin{equation}\label{delrpla1}
  \Delta R = \Delta R_{10} + \Delta R_{12}
\end{equation}
and for planet $P_2$:
\begin{equation}\label{delrpla2}
  \Delta R = \Delta R_{20} + \Delta R_{21}
\end{equation}
where

\begin{equation}
\label{dr10}
     \Delta R_{10} = \nabla_1 R_{10} \Delta \vec{r_1} +  \nabla_0 R_{10} \Delta \vec{r_0}
\end{equation}
\begin{equation}
\label{dr12}
     \Delta R_{12} = \nabla_1 R_{12} \Delta \vec{r_1} +  \nabla_2 R_{12} \Delta \vec{r_2}
\end{equation}
and

\begin{equation}
\label{dr20}
     \Delta R_{20} = \nabla_2 R_{20} \Delta \vec{r_2} +  \nabla_0 R_{20} \Delta \vec{r_0}
\end{equation}
\begin{equation}
\label{dr21}
     \Delta R_{21} = \nabla_2 R_{21} \Delta \vec{r_2} +  \nabla_1 R_{21} \Delta \vec{r_1}
\end{equation}

We finally obtain the three strengths $S_0, S_1, S_2$ for the three planets:
\begin{equation}
\label{sri}
     S_i = (\rho_{max} - \rho_{min})/2
\end{equation}
The strengths $S_i$ as defined above must have some relation, not necessarily linear, with the dynamical effects of the resonance on $P_i$, for example, the width of
the resonance or the amplitude $\Delta a_i$ of the librations observed in the semimajor axis of $P_i$.
The code for this algorithm can be downloaded from www.fisica.edu.uy/$\sim$gallardo/atlas.

\end{document}